\documentclass[iop]{emulateapj}
\usepackage{amsmath}
\usepackage[]{natbib}
\usepackage[usenames]{color} 

\newcommand{\kms}{\>{\rm km}\,{\rm s}^{-1}}

\newcommand{\etal}{{et al.~}}
\newcommand\eg{{\it e.g.}}

\newcommand{\VMons}{V$838$ Mon's }
\newcommand{\VMon}{V$838$ Mon }
\newcommand{\VMonns}{V$838$ Mon}
\newcommand{\halpha}{H$\alpha$ }

\def\apj{{ApJ}}
\def\apjs{{ApJS}}

\def\aj{{AJ}}
\def\mnras{{MNRAS}}

\shorttitle{V838 Monocerotis}

\shortauthors{Loebman \etal}
\begin{document}

\title{The Continued Optical to Mid-IR Evolution of V838 Monocerotis $^{\dagger}$}
\author{S. R. Loebman\altaffilmark{1,2}}
\author{J. P. Wisniewski\altaffilmark{3,4}}
\author{S. J. Schmidt\altaffilmark{5}}
\author{A. F. Kowalski\altaffilmark{6}}
\author{R. K. Barry\altaffilmark{7}}
\author{K. S. Bjorkman\altaffilmark{4,8}}
\author{H. B. Hammel\altaffilmark{9}}
\author{S. L. Hawley\altaffilmark{10}}
\author{L. Hebb\altaffilmark{11}}
\author{M. M. Kasliwal\altaffilmark{12,13}}
\author{D. K. Lynch\altaffilmark{4,14}}
\author{R. W. Russell\altaffilmark{4,14}}
\author{M. L. Sitko\altaffilmark{4,15}}
\author{P. Szkody\altaffilmark{10}}

\altaffiltext{1}{{Department of Astronomy, University of Michigan,
                  830 Dennison,  
                  500 Church Street, Ann Arbor, MI 48109-1042, USA};       
		 {\tt sloebman@umich.edu}}
\altaffiltext{2}{Postdoctoral Fellow, Michigan Society of Fellows}
\altaffiltext{3}{Homer~L. Dodge Department of Physics \& Astronomy, 
                 The University of Oklahoma, 440 W. Brooks Street, 
                 Norman, OK 73019, USA}
\altaffiltext{4}{Visiting Astronomer at the Infrared Telescope Facility, which is operated by the University of Hawaii under Cooperative Agreement no. NNX-08AE38A with the National Aeronautics and Space Administration, Science Mission Directorate, Planetary Astronomy Program.}
\altaffiltext{5}{Department of Astronomy, Ohio State University, 
                 140 West 18th Avenue, Columbus, OH 43210, USA}
\altaffiltext{6}{NASA Postdoctoral Program Fellow, 
                 NASA Goddard Space Flight Center, 
                 Code 671, Greenbelt, MD 20771, USA}
\altaffiltext{7}{NASA Goddard Space Flight Center, 
                 Laboratory for Exoplanets \& Stellar Astrophysics,
                 Code 667, Greenbelt, MD 20771, USA}
\altaffiltext{8}{Ritter Observatory, MS \#113, 
                 Department of Physics \& Astronomy, 
                 University of Toledo, 
                 Toledo, OH 43606-3390, USA}
\altaffiltext{9}{AURA, 1212 New York Avenue NW, 
                 Suite 450, Washington, DC 20005, USA}
\altaffiltext{10}{Department of Astronomy, University of Washington, 
                 Box 351580, Seattle, WA 98195, USA}
\altaffiltext{11}{Department of Physics, Hobart \& William Smith Colleges,
                 300 Pulteney Street, Geneva, NY 14456, USA}
\altaffiltext{12}{The Observatories, Carnegie Institution for Science, 
                  813 Santa Barbara Street, Pasadena, CA 91101, USA}
\altaffiltext{13}{Hubble Fellow}
\altaffiltext{14}{The Aerospace Corporation, M2-266, 
                  P.~O.~Box 92957, Los Angeles, CA 90009-29257, USA ;
                  Guest Observer, NASA Infrared Telescope Facility.}
\altaffiltext{15}{Department of Physics, University of Cincinnati, 
                  Cincinnati OH 45221, USA ; 
                  Guest Observer, NASA Infrared Telescope Facility. ; 
                  Also at Space Science Institute, 4750 Walnut Avenue, 
                  Suite 205, Boulder, CO 80301, USA}
\footnotetext[$\dagger$]{This publication is partially based on observations obtained with the Apache Point Observatory 3.5 m telescope, which is owned and operated by the Astrophysical Research Consortium.}

\begin{abstract} 
The eruptive variable V838 Monocerotis gained notoriety in 2002 when it brightened nine magnitudes in a series of three outbursts and then rapidly evolved into an extremely cool supergiant.   
We present optical, near-IR, and mid-IR spectroscopic and photometric observations of V838 Monocerotis obtained between 2008 and 2012 at the Apache Point Observatory $3.5$m, NASA IRTF 3m, and Gemini South 8m telescopes.  
We contemporaneously analyze the optical \& IR spectroscopic properties of V838 Monocerotis to arrive at a revised spectral type L3 supergiant and effective temperature T$_{\rm eff}\sim2000$--$2200$ K.
Because there are no existing optical observational data for L supergiants in the optical, we speculate that V838 Monocerotis may represent the prototype for L supergiants in this wavelength regime.
We find a low level of \halpha emission present in the system, consistent with interaction between V838 Monocerotis and its B3V binary; however, we cannot rule out a stellar collision as the genesis event, which could result in the observed \halpha activity.   
Based upon a two-component blackbody fit to all wavelengths of our data, we conclude that, as of 2009, a shell of ejecta surrounded V838 Monocerotis at a radius of R$=$263$\pm$10 AU with a temperature of T$=$285$\pm2$ K.  
This result is consistent with IR interferometric observations from the same era and predictions from the Lynch et al.~ model of the expanding system, which provides a simple framework for understanding this complicated system.  
\end{abstract}


\section{Introduction}
\label{sec:intro}
V838 Monocerotis (henceforth \VMonns) is a dramatic stellar object: in 2002, it brightened nine magnitudes during three outbursts \citep{Brown2002,Goranskii2002}, experienced a variety of distinctive optical and infrared (IR) changes \citep{Wisniewski2003,Lynch2004,Rushton2005,Geballe2007}, and eventually morphed into a L-type supergiant \citep{Evans2003}.  
V838 Mon is estimated to be located at a distance of 6.2$\pm$1.2kpc \citep{Sparks2008} and is embedded in a sparse, young cluster with an upper age limit of $\sim$25 Myr and an intervening reddening E(B-V)=0.85 \citep{Afsar2007, Tylenda2012}.  
Since its outburst, an unresolved B3V companion in the system has been detected through spectroscopic monitoring \citep{Munari2005}.

Many theoretical mechanisms have been explored to explain the origin of \VMons outbursts \citep{Lawlor2005,Munari2005,Retter2006,Tylenda2006}; the most likely scenario is a stellar merger between two progenitor stars in a formerly triple system \citep{Tylenda2006}.
This scenerio also predicts x-ray emission from the spin-up of the envelope produced by such a merger \citep{Soker2007}; however, such x-ray emission has not been seen in two epochs of Chandra imagery \citep{Antonini2010}.  
The light from \VMons B3V companion is sufficient to account for the entire luminosity of the variable star measured on sky-survey photographs before its outburst \citep{Afsar2007} \citep[however, see ][for a discussion of the progenitor as its own B3V star]{Barsukova2010}.
  
While \VMon is a decidedly rare object, several analogs have been reported in the literature. 
Two emerging classes of cool explosions are: luminous red novae (LRNe) and intermediate luminosity red transients (ILRTs). 
Akin to \VMonns, LRNe are stellar eruptions that have remained extremely cool through the outburst and include V1309 Sco \citep{Tylenda2011} and V4332 Sgr \citep{Martini1999}. 
Two extragalactic events have also been noted to be LRNe: one in the bulge of M31 \citep{Mould1990} and one in the lenticular galaxy M85 \citep{Kulkarni2007}. 
It has been proposed that LRNe are stellar mergers. 
ILRTs are similarly red but too luminous to be explained by stellar mergers: \eg, SN2008S in NGC 6946 \citep{Prieto2008}, NGC300-OT2008 \citep{Bond2009, Thompson2009}, and PTF10fqs in M99 \citep{Kasliwal2011}. 
Consistent with their discovery in grand spirals, it has been proposed that ILRTs represent electron-capture induced collapse in extreme Asymptotic Giant Brach Stars \citep{Kochanek2011}.

The local circumstellar environment of \VMon has been studied extensively since the initial outburst of the system.  
Based on an observed, short-lived (February -- March, 2002) intrinsic polarization in the system, the outburst geometry was likely non-spherical \citep{Wisniewski2003}.
In October 2002, a temporary re-emergence of an intrinsic polarization component occurred.
Oriented 90$^{\circ}$ from the original component, the new polarization suggests one of two changes to the system: either a change in the optical depth of the material surrounding the star(s) or a physical shift in the illuminating source(s) \citep{Wisniewski2003b}.  
Recent interferometric observations \citep{Chesneau2014} have indicated a similar local circumstellar geometry as suggested by polarimetric data.  
V838 Mon is also surrounded by a substantially broader region of nebular material as diagnosed by the dramatic light echo illuminated by the outburst events \citep{Bond2003}.
 
Continued observations of V838 Mon's local circumstellar environment have traced the dynamical evolution of the system.  
Optical and IR photometric monitoring has suggested that the outburst ejecta has advanced past the location of the B3V binary, completely attenuating its signal \citep{Goranskiji2008} and condensing to form new circumstellar dust \citep{Wisniewski2008}. 
To explain the evolution of spectroscopic and photometric properties observed over time, \citet{Lynch2004} proposed a simple, spherically symmetric model containing a central star with an expanding circumstellar shell that is cooling in a radiatively-dominated quasi-equilibrium manner; this model is consistent with spectral fits to optical and IR data from \citet{Lynch2004}. 
More recently, \citet{Lynch2007} expanded this conceptual framework to include five components: a central star, two photospheric shells that surround the central star, and two much cooler and more distant expanding shells of gas.  
Validating the central tenents of these models, including the number of components required and their expansion velocities, has remained elusive due to the previously limited number of epochs (2) of IR data available.
 
In this paper, we discuss the continued evolution of the system in the context of the Lynch et al.~ (2004, 2007) models and previously published photometric, spectroscopic, and spectro-polarimetric observations of the system. 
In \S\ref{sec:observ}, we present optical, near-IR, and mid-IR spectroscopic and mid-IR photometric observations of \VMon obtained between 2008 and 2012 at the Apache Point Observatory 3.5m, NASA IRTF 3m, and Gemini South 8m telescopes.
In \S\ref{sec:halpha}, we show a low level of \halpha emission has returned to the system, possibly due to excitation of the gas by the B3V companion as seen through the optically thinning gas and dust ejecta or due to magnetic activity produced by an emerging dynamo effect from the proposed stellar merger.
In \S\ref{sec:spectyp}, we spectrally type \VMon in both the optical (M7 supergiant) and near infrared (L3 supergiant) and discuss the discrepancy in the two identifications.
In \S\ref{sec:irseds}, we fit our 2009 observations to the \citet{Lynch2004} model for the expanding warm dust envelope and find our data is consistent with the previous two epochs of data.
Finally, in \S\ref{sec:results}, we discuss the implications of the results we presented in the previous sections.

\section{Observations}
\label{sec:observ}

Optical spectra were taken using the Astrophysical Research Consortium Echelle Spectrograph (ARCES) and the Dual Imaging Spectrograph (DIS) instruments on Apache Point Observatory's (APO) $3.5$m telescope between 2008 - 2012 (see Table 
\ref{tab:observation_table}).
ARCES \citep{Wang2003} is a high resolution, cross-dispersed visible light spectrograph \footnote[16]{http://www.apo.nmsu.edu/arc35m/Instruments/ARCES/} that obtains spectra between 3600-10000 $\AA$ with a resolution of R$\sim$31500.  Standard bias, flat field, and ThAr lamp exposures for the echelle were also obtained on every night.  
These data were reduced using standard techniques in IRAF\footnote[17]{IRAF is distributed by the National Optical Astronomy Observatories, which are operated by the Association of Universities for Research in Astronomy, Inc., under cooperative agreement with the National Science Foundation.}.  

DIS is a medium dispersion spectrograph with separate red and blue channels.\footnote[18]{http://www.apo.nmsu.edu/arc35m/Instruments/DIS/}
Our DIS observations were taken using the default low-resolution blue and red 
gratings (B400/R300), yielding a resolution R$\sim$800 and a total effective 
wavelength coverage of $4500-9000 \AA$.  Bias, flat field, and HeNeAr lamp exposures for DIS were obtained on every night, 
and nightly observations of the flux standard stars G191B2b, Feige34 or Feige110 were used to flux calibrate these data.  
These data were reduced using standard techniques in IRAF.

We obtained two epochs of observations of V838 Mon 
(Table \ref{tab:observation_table}) using SpeX \citep{Rayner2003}, a medium-resolution $0.8$-$5.4$ $\mu$m spectrograph located at NASA's Infrared Telescope Facility (IRTF).\footnote[19]{http://irtfweb.ifa.hawaii.edu/$\sim$spex/}  We used
the 0$\farcs$3 x 15$\farcs$0 slit for our V838 Mon observations, providing R$\sim$2000 spectra from 0.8-2.5 $\mu$m, and also 
obtained observations of the A0V star HD 53205 to provide telluric correction.  These data were reduced using 
Spextool \citep{Vacca2003,Cushing2004}.

Mid-IR low resolution spectroscopy and imaging of V838 Mon were obtained with
T-ReCS (Thermal-Region Camera Spectrograph), located at the Gemini South Observatory (GSO).\footnote[20]{http://www.gemini.edu/sciops/instruments/trec/}  As summarized in Table \ref{tab:observation_table}, we 
obtained both low resolution (R$\sim$100) spectroscopy in the N-band, narrow-band photometry near the N-band 
(Si-5: 11.7 $\mu$m and Si-6: 12.3 $\mu$m), and broad-band photomtry in the Q-band (Qa: 18.3 $\mu$m).  Standard telescope 
chopping and nodding was used to mitigate the effects of the thermal background of the sky in the mid IR.  These 
data were reduced using the \textit{midir} Gemini/IRAF software package.

We also present two epochs of mid-IR photometry of V838 Mon obtained with the Broadband Array Spectrograph System (BASS), mounted on NASA's IRTF (see Table \ref{tab:observation_table}).  BASS is an infrared array prism spectrograph \citep{Hackwell1990} that covers the $2.9$-$13.5$ $\mu$m spectral region simultaneously at a resolving power of 25-120, depending on wavelength, and records these data on 116 back-illuminated blocked impurity band Si:As detectors.\footnote[21]{http://www.aero.org/capabilities/remotesensing/bass.html} Our BASS observations were calibrated using $\alpha$ Tau as a reference star. We note that relative calibrations for reference stars $\alpha$ Tau, $\alpha$ CMa, $\beta$ Gem, $\alpha$ Boo, and $\alpha$ Lyr have remained constant to about 1\% with repeated tests.

\section{\halpha Evolution}
\label{sec:halpha}

The \halpha line has proven to be an important diagnostic of the evolution of V838 Mon's circumstellar environment, helping 
to trace the initial expansion of ejecta from the 2002 outbursts \citep{Wisniewski2003} and in 2006 probing the likely interaction of this expanding ejecta with the B3V companion \citep{Munari2007}.  We discuss the continued monitoring of 
 \halpha between October 2008 and September 2012 below.

\begin{figure*}[!H]
\center{\includegraphics[width=1\textwidth]{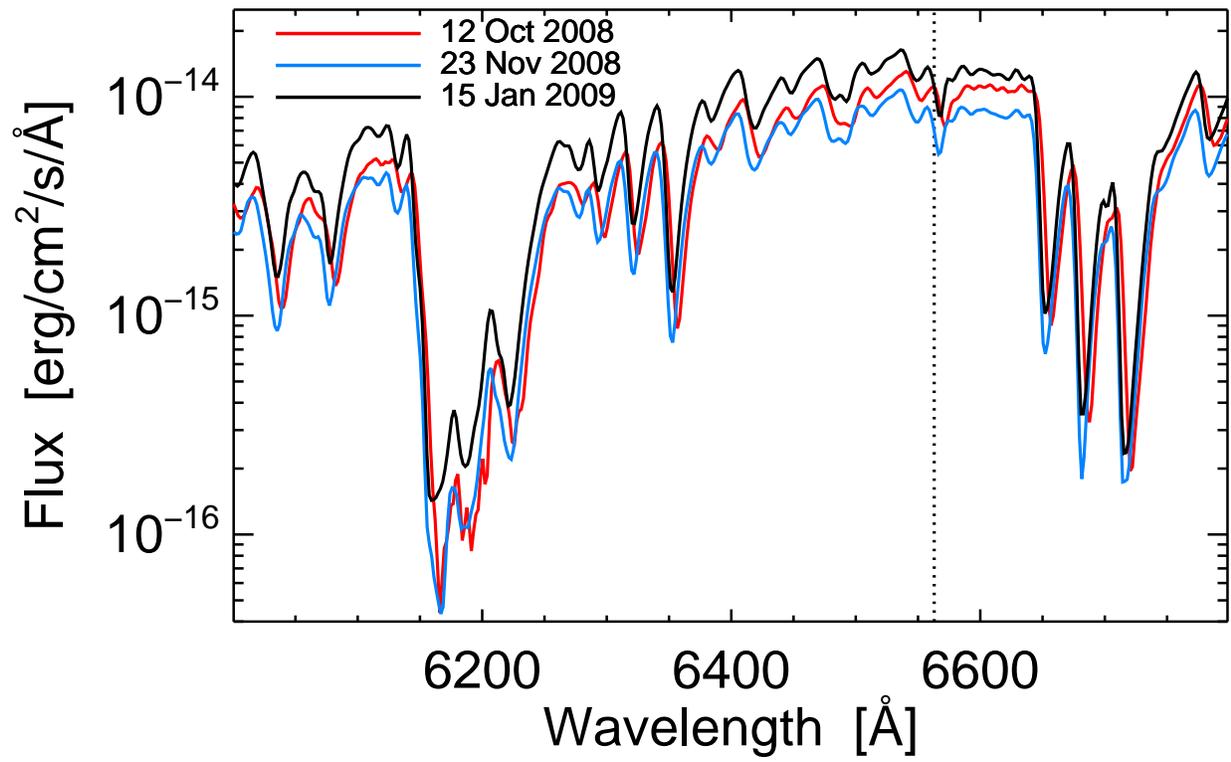}}
\caption{We monitored the optical spectrum of \VMon at low resolution (R $\sim 800$) with the DIS instrument on the $3.5$m Apache Point Observatory telescope from 2008$-$2009.  We find no convincing evidence of emission at this resolution, consistent with that noted by \citet{Bond2009b}.}
\label{f:DIS_halpha}
\end{figure*}

\begin{figure*}[!H]
\center{\includegraphics[width=1\textwidth]{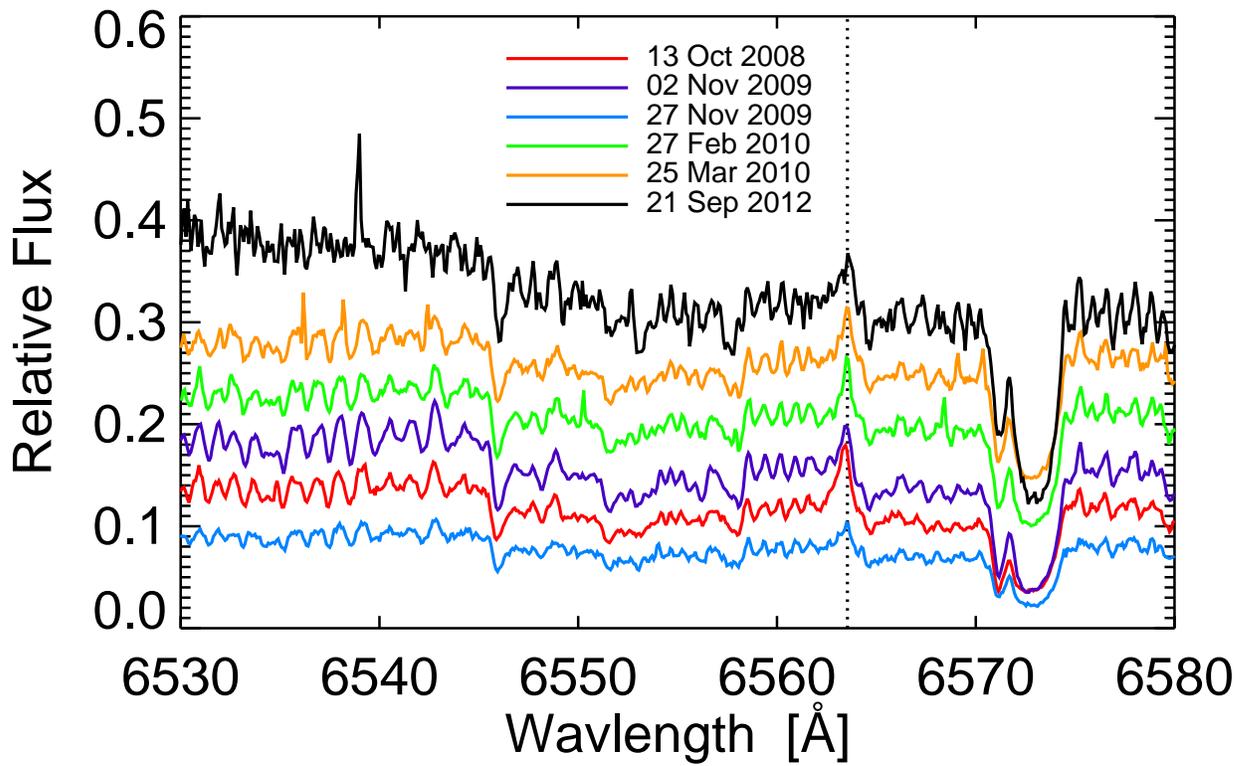}}
\caption{We also monitored \VMon using the high resolution (R $\sim 31500$) Echelle spectrograph on the $3.5$m Apache Point Observatory telescope.  At this resolution, we see clear evidence of a low level of \halpha emission from October 2008 $-$ September 2012.  Note, we have added a fiducial offset to each spectrum for ease of comparison.}
\label{f:ECHELLE_halpha}
\end{figure*}

In Figure~\ref{f:DIS_halpha}, we present three optical spectra of \VMon taken with the low resolution (R $\sim 800$) DIS spectrograph on APO.  We detect no convincing evidence of \halpha emission at this resolution, which 
is consistent with conclusions drawn by \citet{Bond2009b}.  However, the three optical spectra of \VMon taken with the high resolution (R $\sim 31500$) Echelle spectrograph on APO during the same epoch as our low resolution spectra clearly 
reveal the presence of a low level of \halpha emission (Figure~\ref{f:ECHELLE_halpha}).  This low level of emission still 
appears present during our fourth epoch of high resolution spectra obtained in 2012.  We do not observe any significant variability in the \halpha emission.  These results are consistent 
with the report of low level \halpha emission being present in 2009-epoch high resolution spectra by \citet{Tylenda2011}, 
and extend the time period over which the system exhibits this emission.

One possible origin for this low level emission could be excitation of the expanding dust and gas envelope by the B3V companion.  \citet{Bond2009b} suggests that the behavior of \halpha from 200 to 2009, which included a brief return of the line appearing in emission followed by a return to an apparent pure absorption profile, was due to the B3V companion becoming completely engulfed by the expanding envelope.  However, our high resolution spectra suggest that perhaps this envelope
is not completely optically thick.  Alternatively, the observed low level \halpha emission could be caused by magnetic activity in the primary star, which is an expected byproduct of the proposed stellar merger event of 2002 \citep{Soker2007}.  We discuss the plausibility of each of these scenarios in \S\ref{sec:results}.

\section{Spectral Typing}
\label{sec:spectyp}

To better understand the properties of the central star, we compare both the DIS optical spectrum and SpeX IR spectrum with other cool objects. For the purposes of this discussion, we assume that the dust shell has little or no effect on the spectral features, and have corrected for a reddening of $E{(B-V)} = 0.85$.

\subsection{Optical Spectral Type}
\label{sec:opttype}
\begin{figure*}[!H]
\center{\includegraphics[width=1\textwidth]{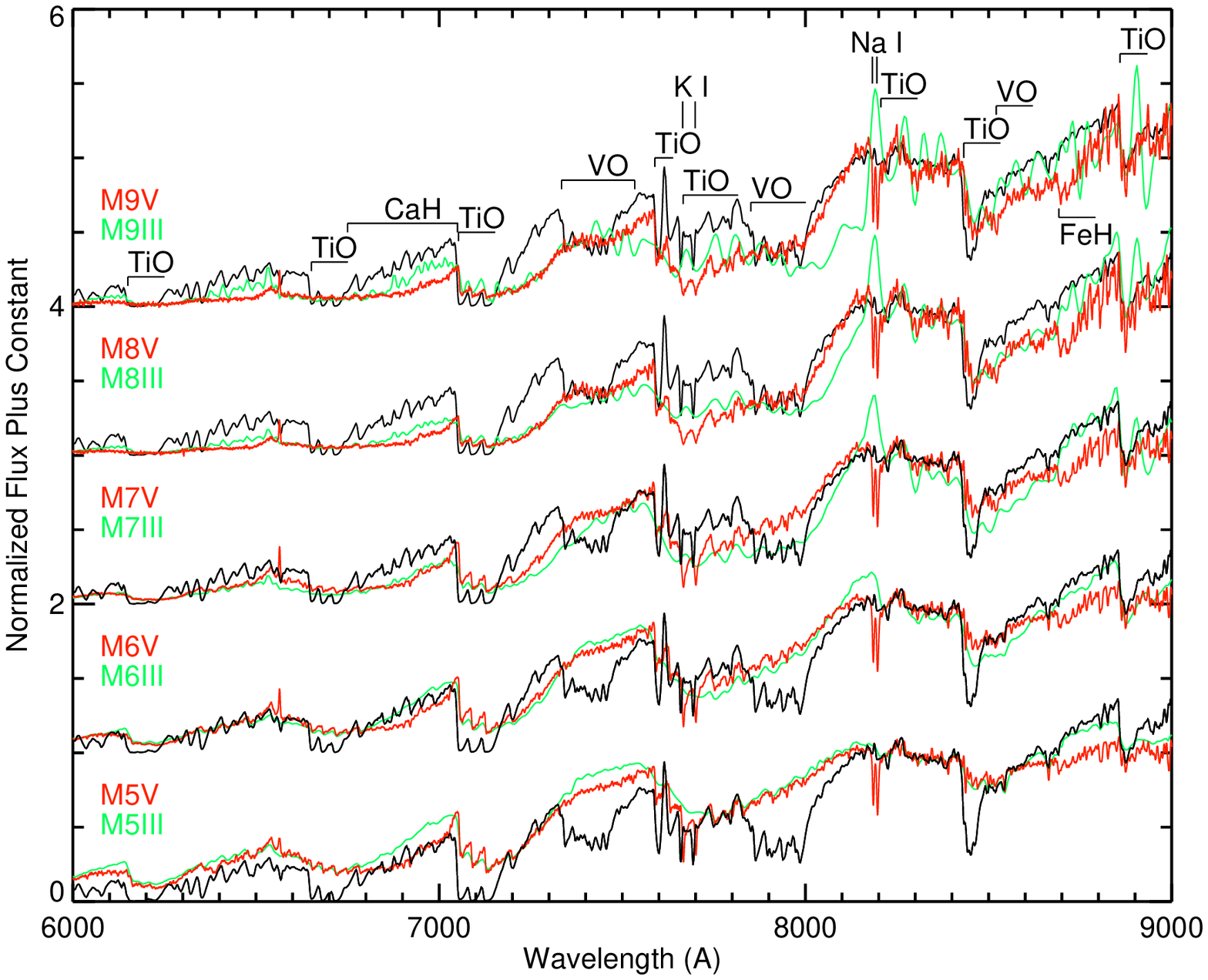}}
\caption{Comparison of our low-resolution ($R \sim 800$), extinction corrected ($E{(B-V)} = 0.85$) optical spectrum of \VMon (black) with M dwarf templates (red) from \citet{Bochanski2007} and M giant templates (green) from \citet{Pickles1998}. Prominent molecular and atomic features are labeled. Due to the dramatic TiO bands, the best match spectra are M7. We note that the optical spectrum of \VMon shows narrower molecular bands and atomic lines than both the dwarfs and giants due to its low surface gravity.}
\label{f:optical_comp}
\end{figure*}

For a starting point in our analysis, we compare the optical spectrum of \VMon to M5--M9 dwarf \citep{Bochanski2007} and giant \citep{Pickles1998} spectroscopic templates (shown in Figure~\ref{f:optical_comp}). We used the Hammer spectral typing software \citep{Covey2007} to estimate the optical spectral type based on the characteristics of M dwarf optical spectra. We obtain a best fit type of M7, consistent with the \citet{Tylenda2011} type of M6.3. An examination of the specific optical features, however, indicates that the type is primarily based on the strength of the TiO bands, and most optical features are more sensitive to the \VMons surface gravity rather than its surface temperature. 

The optical spectrum of \VMon shows strong, sharp TiO and VO features, with only weak CaH absorption at 6750\AA~and no sign of FeH absorption at 8600\AA. The presence and strength of TiO and VO is consistent with an oxygen-rich stellar atmosphere, and the absence of FeH and CaH bands indicate a very low gravity atmosphere; those bands are weak or absent in M giants \citep{Evans2003}. The sharpness of the TiO and VO bands and the \ion{K}{1} doublet at 7700\AA~compared to both the dwarf and giant templates indicates that it is likely to have an even lower surface gravity (consistent with a supergiant classification). The TiO bands at 6150\AA, 6651\AA, and 7053\AA~each absorb to zero flux, indicating the absence of the B3 component detected by \citet{Munari2005}. 

There is no detection of the 8183/8195\AA~\ion{Na}{1} doublet, which is typically in absorption for late-M dwarfs and in emission for late-M giants \citep{Schiavon1997}. It is possible that the lack of emission is due to the lower gravity of \VMon compared to M giants. The absence of this feature is perhaps not remarkable due to the weak or absent \ion{Na}{1} absorption both in the optical \citep[the $\sim$5893\AA~doublet is very weak in the 2009 spectrum]{Tylenda2011} and the near-infrared (we do not detect the \ion{Na}{1} doublets at $\sim$1.14 and $\sim$2.21$\mu$m; see Section~\ref{sec:irtype}). 

\subsection{Infrared Spectral Type}
\label{sec:irtype}
\begin{figure*}[!H]
\center{\includegraphics[width=1\textwidth]{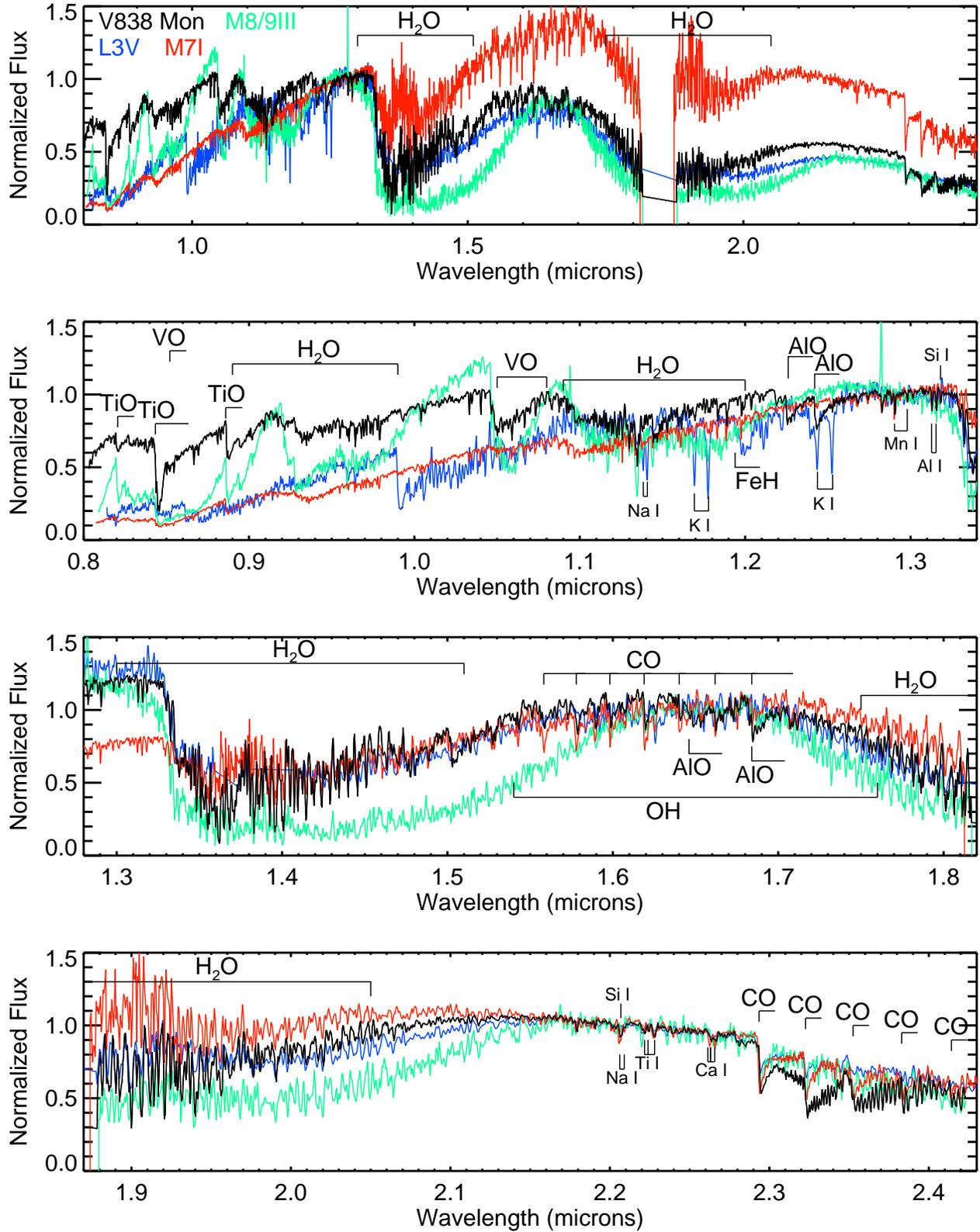}}
\caption{Comparison of our IRTF SpeX data (black lines) to spectra from the IRTF Spectral Library \citep[colored lines][]{Cushing2005a,Rayner2009}. The top panel shows the entire range of the SpeX data, while the next three panels show the $J$-, $H$- and $K$-band respectively. The types of each spectrum are given in the top left corner of the top panel. None of the comparison spectra are a perfect match for the \VMon spectrum. The L3V is the best match for the overall spectral slope, while the M8/9III is the best match in the $J$-band due to the dramatic TiO bands. The M7I spectrum is the best match in the $H$- and $K$-bands.}
\label{f:ir_comp}
\end{figure*}

\citet{Evans2003} noted that the 0.8--2.4$\mu$m spectrum of \VMon appeared to be that of an L supergiant star. There are no other objects with an L supergiant classification, so we select a range of spectra for comparison to the SpeX observations of \VMon, drawing both from late-M and L dwarf spectra and late-M giant and supergiant spectra. In Figure~\ref{f:ir_comp}, we show the SpeX spectrum of \VMon compared to publicly available infrared spectra \citep{Cushing2005a,Rayner2009} of M8V LP 412-31 (similar to its initial optical classification), M8/9III IRAS 14303-1042 (as an example late-M giant), and M7I MY Cep (as an example of a late-M supergaint), and L3V 2MASS J00361617+1821104 (possibly similar in surface temperature) from the IRTF Spectral Library\footnote[22]{http://irtfweb.ifa.hawaii.edu/\~spex/IRTF\_Spectral\_Library/index.html}. Each of these spectra share some spectral features with \VMon. 

The best match for the overall shape of the \VMon spectrum is a mid-L dwarf, so we apply L dwarf classification indices to the infrared spectrum of \VMon. L dwarfs are classified not only on their spectral type/T$_{\rm eff}$, but also their surface gravity \citep[which shows a dependency on age;][]{Burrows1997}. Based on the strength of its H$_2$O bands, which primarily trace effective temperature, \VMon has an infrared spectral type of L3. The \citet{Allers2013} gravity indices (based on the strength of \ion{K}{1}, VO, and FeH) indicate a very low gravity. The \ion{K}{1} and FeH are in fact completely absent from the spectrum, consistent with the very low surface gravity L supergiant classification \citet{Evans2003}. 

Despite the L3 classification, the detailed spectral features of \VMon are not a close match to those of L3 dwarfs, as can be seen in Figure~\ref{f:ir_comp}. The TiO bands between 0.8 and 0.9 microns, discussed in \ref{sec:opttype}, are not typically present in L dwarf spectra. The very low surface gravity of \VMon results in the absence of FeH, \ion{K}{1} and \ion{Na}{1} between 1.1 and 1.3$\mu$m. Another dramatic difference between L dwarfs an \VMon is the presence of AlO A-X absorption bands \citep[first identified on \VMon by][]{Bernstein2003} at 1.226, 1.242, 1.646, and 1.648$\mu$m. These bands are consistent with oxygen-rich, low temperature environments and have been observed in a handful of AGB stars \citep{Banerjee2012}.

The best match for the detailed $J$-band spectrum of \VMon is that of the M8/9II IRAS 14303-1042; both have TiO, VO, and H$_2$O absorption features, though the specific shapes and strengths of the features are not well matched. The $H$-band spectrum of \VMon is dominated by H$_2$O, CO, and OH, resulting in detailed features very similar to those of the M7I MY Cep (excepting the AlO absorption bands, present only in \VMon). The $K$-band spectra of all three comparison stars are a good overall match to \VMon, but the CO bands appear different in both shape and strength, possibly due to the complexities of the dust shell surrounding \VMon. 

The CO bands in our 2008 spectrum are qualitatively similar to those detected in the \citet{Geballe2007} low-resolution 2006 infrared spectra. \citet{Geballe2007} analyzes the CO bands in detail using a high resolution (R$\sim$18000) spectra taken within a few months of the low resolution detection; this high-resolution spectra revealed distinct velocity components within the CO spectrum. The main component of the CO was at photospheric temperatures, but with a radial velocity of $-15$~km~s$^{-1}$ relative to the stellar radial velocity, indicating the gas was still settling on the surface of the star. Other components of the CO absorption were associated with the more extended dust shell. In our 2008 spectrum, \VMons CO bands are stronger than those of all the comparison objects, which may indicate that some of the extended dust shell components still contribute to the absorption, but we do not have the velocity resolution to investigate the CO bands in more detail. 

\VMon is relatively free of molecular absorption near 1.3 and 2.2$\mu$m, revealing weak absorption lines. These lines appear qualitatively similar to those of the M7I MY Cep and are likely a feature of supergiant atmospheres. Using the list of infrared lines in Arcturus \citep{Rayner2009}, we tentatively identify a \ion{Mn}{1} $\lambda$ 1.2903, 1.2980$\mu$m doublet; an \ion{Al}{1} $\lambda$1.3127, 1.3153$\mu$m doublet; and a \ion{Si}{1} $\lambda$1.3181$\mu$m absorption line in the 1.25--1.32$\mu$m spectral region. Near $\sim$2.2$\mu$m, we tentatively identify \ion{Si}{1} at 2.2069$\mu$m (not the \ion{Na}{1} doublet seen in the dwarf spectra); a \ion{Ti}{1} $\lambda$2.2217, 2.2239, 2.2280$\mu$m triplet; and a \ion{Ca}{1} $\lambda$2.2614, 2.2631, 2.2633, 2.2657$\mu$m quadruplet. Both regions have additional lines of similar strength that overlapped with multiple atomic transitions so could not be positively identified at this resolution. 

\subsection{Estimating T$_{\rm eff}$ From Spectra}
The optical and infrared spectra of \VMon, at first glance, seem to indicate different spectral types, and therefore different surface temperatures (T$_{\rm eff}$). Adopting the M7 optical spectral type and extrapolating the supergiant spectral type/T$_{\rm eff}$ relation of \citet{Levesque2005} gives a surface temperature of T$_{\rm eff}\sim3000$~K \citep[similar to the temperature quoted by][]{Tylenda2011}. There is no supergiant temperature scale for L dwarfs, but an L3 dwarf spectral type corresponds to T$_{\rm eff}\sim1800$~K \citep{Stephens2009}. One possible way of reconciling those two temperatures is an application of a multi-layer photosphere \citep[e.g.,][]{Lynch2007}, but here we estimate based on the comparison of \VMon to other stars. 

As discussed in \S~\ref{sec:opttype}, the optical spectrum for \VMon matches an M7 dwarf best because the TiO bands are strongest on M7 photospheres. The fraction of TiO contained in molecules (vs.~ \ion{Ti}{1}) increases with decreasing temperature, hitting a maximum (at M dwarf atmospheric pressures) at T$_{\rm eff}\sim2000$~K \citep{Lodders2006}. At temperatures cooler than T$_{\rm eff}\lesssim2400$~K, TiO begins to condense onto dust grains (e.g., perovskite) \citep{Burrows1999}, so TiO disappears in M7 and later dwarfs. The formation of dust grains is more sensitive to atmospheric pressure than the formation of TiO, so at low surface gravity, a low temperature atmosphere will continue to have strong TiO bands at T$_{\rm eff}\lesssim2400$~K. 

The remaining indicators of cool temperatures in optical spectra (e.g., the transition from late-M to L dwarfs) are pressure sensitive; gravity broadening of the \ion{K}{1} doublet, the strengthening of FeH. Additionally, the infrared spectrum of \VMon appears cooler than the M7I MY Cep due to stronger TiO, VO, and H$_2$O bands. Therefore, the spectrum of \VMon may be both very cooler and very low gravity; a prototype for L supergiants in the optical. M giants are T$\sim200$--$400$~K warmer than dwarfs of the same optical type, and if this trend continues, the surface temperature of \VMon is likely to be T$_{\rm eff}\sim2000$--$2200$~K. This range of temperatures is also consistent with the formation of strong TiO bands without significant depletion of TiO onto dust. 

\section{Infrared SEDS}
\label{sec:irseds}

In  this section,  we fit  our visible  and infrared  data with  a two
component blackbody curve.  We consider the physical interpretation of
this fit in context  of the \citet[][henceforth, L04]{Lynch2004} model
of \VMons post  outburst evolution. Our goal is to constrain the
evolution of the ejecta with our new epoch of data. 
The L04 model contains three components.
\begin{enumerate}
\item{The central star's photospheric  emission, which is modeled as an
  inner blackbody  source with  a homogeneous outer,  cooler absorbing
  layer. The  central star  is described by  a radius, $R_{p}$,  and a
  blackbody temperature, $T^{p}_{BB}$.}
\item{A  warm  shell,  which   is  modeled  with  several  parameters,
  including an expanding  radius, $R_{s}$, and a vibrational/blackbody
  emission temperature, $T^{s}_{BB}$.}
\item{A cold, outer  shell, modeled in a similar  fashion as the warm,
  expanding  shell, but  with  an emission  temperature  set to  zero.
  Here, the  quantity of greatest significance is  the absorber column
  (resulting in an effective emissivity, $\epsilon$, of the system).}
\end{enumerate}

We also  considered the \citet[][henceforth,  L07]{Lynch2007} model of
\VMons  ejecta,  which contains  5  components  (a  central star,  two
photospheric  shells that  surround  the central  star,  and two  much
cooler  and more  distant expanding  shells  of gas).   This model  is
essentially the  same as  the L04 model,  except the  two photospheric
layers are considered  separately from the central star,  and are each
assumed to  be homogeneous and  characterized by a  single equilibrium
temperature and  column amounts for the molecular  species.  While the
added  components  are  physically  reasonable  and  lead  to  refined
predictions  for  $T^{p}_{BB}$,   the  added  complexity  beyond  this
revision  did  not help  us  build  additional  insight from  our  two
component  blackbody fit.   Therefore, our  remaining  discussion will
focus primarily on comparisons with the L04 model.

Best fits from L04 and L07 can be found in Table~\ref{table:model}.
We note that the generally accepted values  for the velocity of the  expanding cloud in 2003 were within  100-400 km s$^{-1}$ (see L04 and references therein). 
We assume  the velocity of the
expanding cloud is $v_{s}=160$ km  s$^{-1}$ and the distance to \VMon is
$D_{V838  Mon}=6$  kpc \citep{Sparks2008} to generate model predictions.  In January 2009, L04 predicts: $R_{p} \sim 4.2$  AU (no evolution from
2005); $T^{p}_{BB} \sim 2100$  K (no  evolution from  2005); $R_{s}
\sim 230.5$ AU  (time elapse 6 year 10.5  months, expansion rate 33.75
AU/yr); and  $T^{s}_{BB} \sim 283$ K (temperature of cloud in 2003 $\times$ 0.38).

\begin{figure*}[!H]
\center{\includegraphics[width=1\textwidth]{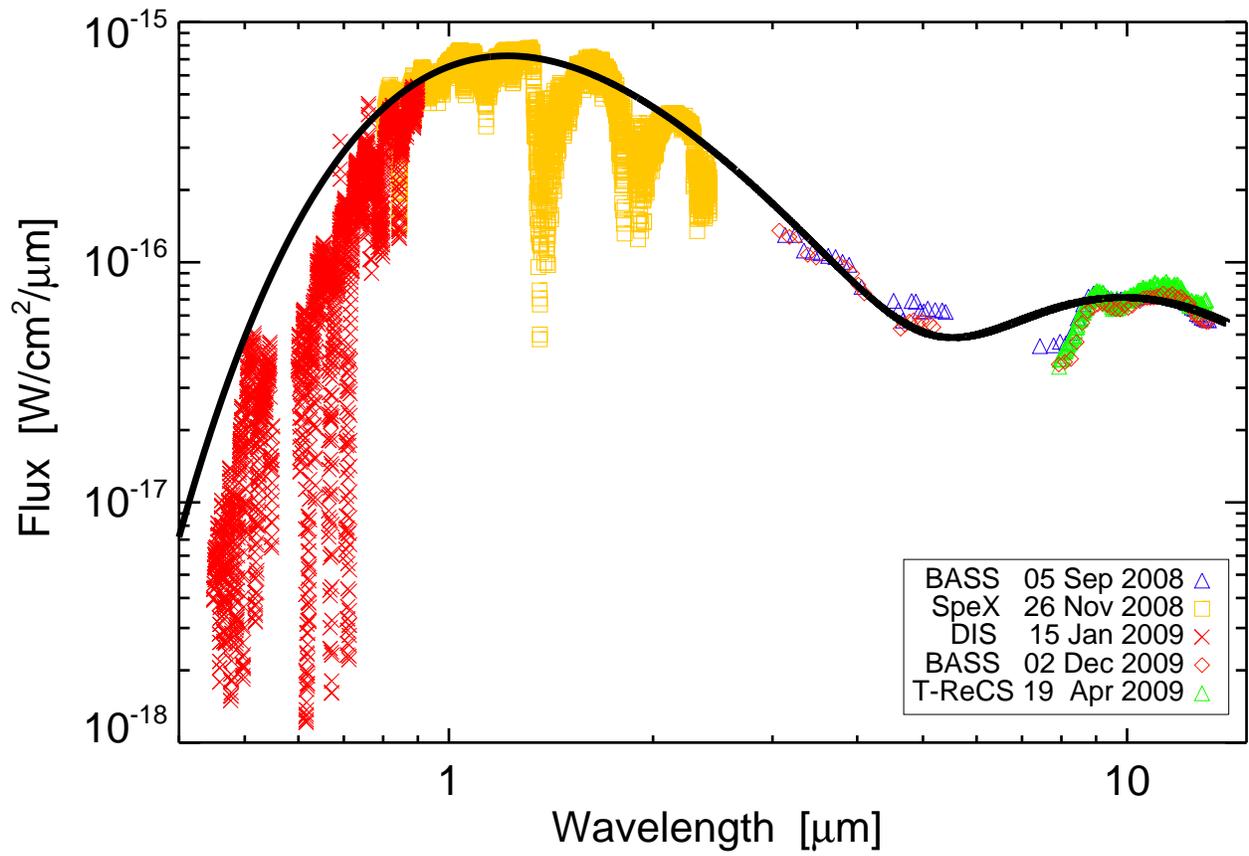}}
\caption{We fit a two component blackbody model to \VMonns, including a 2370 K component from the primary star and a 285 K component from the warm expanding envelope.  These data provide observational support for the \citet{Lynch2004} model.}
\label{f:bb}
\end{figure*}

We fit  our 2009 SED  data, consisting of both  de-reddened optical
and  infrared  data,  with  a  2 component  blackbody  curve (Figure \ref{f:bb}).   Our
simultaneous fit yielded for  the primary star T$_{p}$$=$2370$\pm400$ K and
R$_{p} =  4.3\pm.2$ AU and T$_{s}$$=$285$\pm2$ K,  R$_{s}$$=$263$\pm10$ AU, and
veiling=$35\%\pm20$ for the warm expanding dust envelope.
We note, the T$_{\rm eff}$ identified in section \S\ref{sec:spectyp} using spectral analysis (T$_{\rm eff}\sim2000$--$2200$ K) falls within the 3$\sigma$ errors on our best fit for the temperature of the primary, $T_p=2370\pm400$ K.

Our results are consistent with the predictions from the L04 model.
We find only a slightly larger radius for the expanding dust shell ($R_{s}$ $\sim$ 263 $\pm$10 AU) 
than is predicted ($R_{s}$ $\sim$ 230.5 AU) 
using an expansion rate  of  $v_{s}=160$  km   s$^{-1}$.  Moreover, the blackbody temperature 
we determine for this shell ($T^{s}_{BB}$ $\sim$ 285 $\pm$ 2 K) is consistent 
with that predicted from the L04 model.  
 
\section{Discussion \& Conclusions}
\label{sec:results}

In this section, we discuss and present our conclusions regarding \VMons spectral type, expanding outburst ejecta, and ongoing \halpha emission.  We also comment on the lack of new dust formation (relative to 2008 observations) and place this in the context of the evolution of the system as a whole.

In \S\ref{sec:spectyp}, we concluded that the IR spectral slope of \VMon is best matched to an L3 supergiant.  
While the optical data is best matched by a spectral classification of M7 (as matched to a dwarf template), this classification does not take into account the effects of low gravity on the persistence of various spectral features.  
Significantly, the temperatures at which TiO can remain coherent in supergiant photospheres could be as high as 2000--2200 K.  
Given the complete lack of L supergiant data to type against within the optical regime, we conclude that \VMon could plausibly be catagorized as a L supergiant in both the IR and optical regimes.  
In this regard, we speculate that \VMon could be a prototype for what an L supergiant optical spectra should look like.

In \S\ref{sec:irseds}, we used our 2009 optical to mid-IR observations as a third epoch of data to test the \citet{Lynch2004} model of the expanding ejecta envelope. 
Our data indicate the dust shell has continued to cool at a similar rate as suggested by the \citet{Lynch2004} model.
	If the shell was ejected at a velocity of $160 \kms$, the 2009 IR data is consistent with a 285$\pm$2 K shell of material which has expanded to a radius of 263$\pm$10 AU and is attenuated by $\sim 35\%\pm20$ by an outer cooler shell.  In this best fit to the model, the radius of the L3 primary is described by $R_{p}\sim4.3\pm0.2$ AU and a $T_{p}\sim2370\pm400$ K. We note, the T$_{\rm eff}$ identified in section \S\ref{sec:spectyp} using spectral analysis (T$_{\rm eff}\sim2000$--$2200$ K) falls within the 3$\sigma$ errors on our best fit for the temperature of the primary, $T_p=2370\pm400$ K.

These results are consistent with the recent IR interferometric analysis by \citet{Chesneau2014}.  
Using the Very Large Telescope Interferometer between October 2011 and February 2012, and adopting a distance to \VMon of $D=6.1\pm0.6$ kpc, \citet{Chesneau2014} deduce that there is a dust shell distributed from about 130 to 300 AU around \VMonns.  
They note the dust is distributed in a flattened structure which could be interpreted as a relic of the genesis event or could be influenced by the embedded B3V companion. 
They also note that as of 2014, the radius of \VMons photosphere has decreased by about 40\% from the radius determined after the outbursting events to $3.5\pm0.6$ AU.  
We note that the radius of the primary that we determined ($4.3\pm0.2$ AU) is within errorbars of the radius determined by \citet{Chesneau2014}.

\begin{figure*}[!H]
\center{\includegraphics[width=1\textwidth]{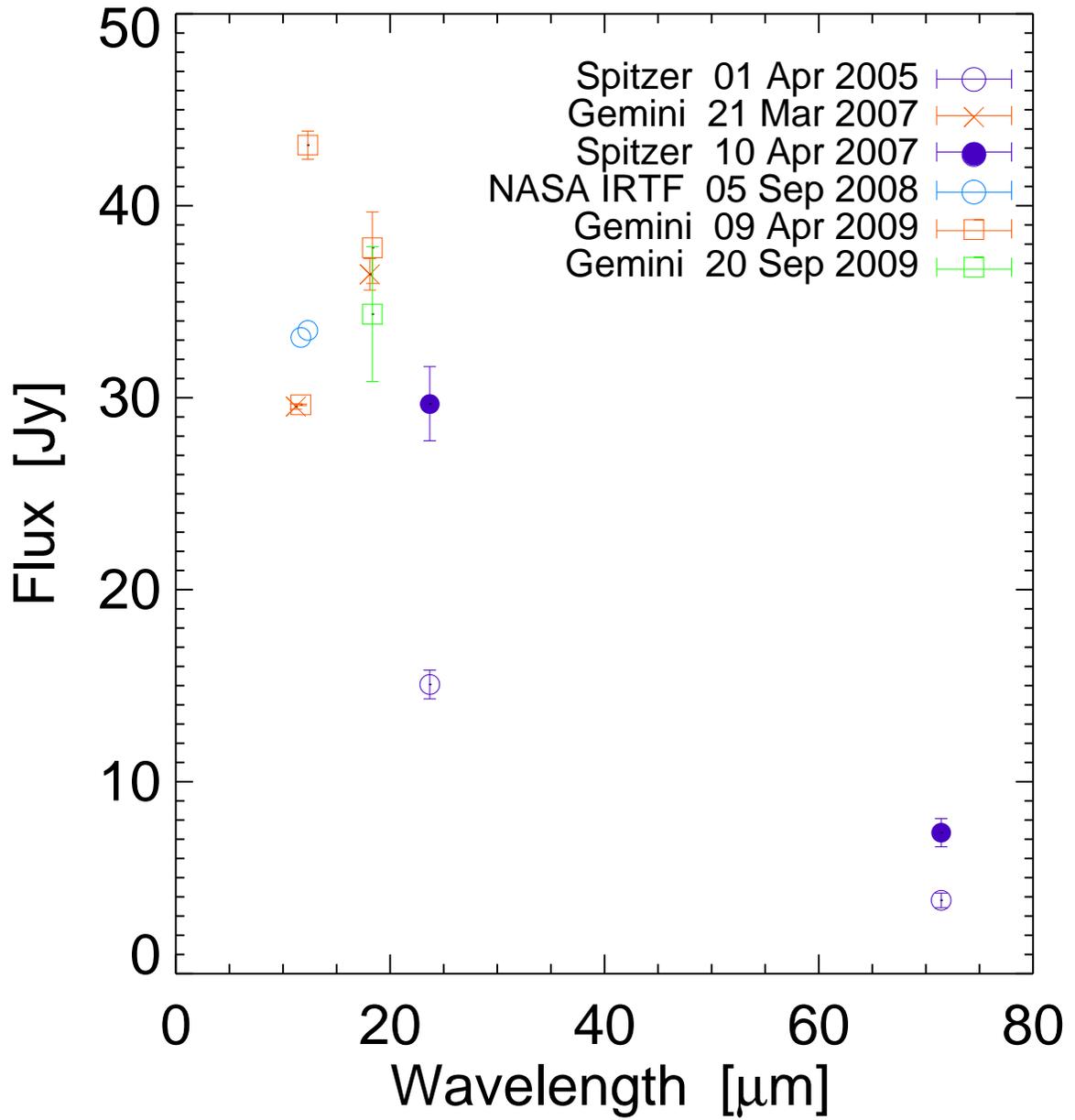}}
\caption{Our 2008$-$2009 IR photometry shows no appreciable change from the 2006$-$2007 epoch \citep{Wisniewski2008}.  We conclude there has been no substantive change in the dust production or composition.}
\label{f:photo}
\end{figure*}

We consider here whether the dust is newly formed material or persisting matter that has expanded to a larger spatial distribution.  
\citet{Wisniewski2008} previously found substantial dust production in the 2006$-$2007 epoch, which was reflected in an IR excess at the 10--80$\mu$m wavelength range. 
We repeat their analysis on our 2008$-$2009 IR photometry.
In Figure~\ref{f:photo}, we plot our new results together with the data presented in \citet{Wisniewski2008}.  
As evident, our 2008$-$2009 IR photometry shows no appreciable change from the 2006$-$2007 epoch.  
We conclude there has been no substantive change in the dust production or composition.

In \S\ref{sec:halpha}, we showed the level of \halpha activity has declined since the line's brief return to significant emission in 2006.   While \halpha emission is not distinguishable in low resolution spectra obtained during 2008-2009, all high-resolution (R $\sim 31500$) spectra obtained between 
2008-2012 do show the ongoing presence of a small amount of \halpha emission.  We suggest two possible origins for the 
observed weak \halpha emission.
The emission could originate from magnetic activity in the primary star, which is an expected phenomenon \citep{Soker2007} if the 2002 outburst was caused by a stellar merger event \citep{Tylenda2006}.  \citet{Antonini2010} found 
highly variable X-ray emission associated with \VMonns: in 2008, strong X-ray emission was detected with XMM Newton, but in 2009 and 2010 no such emission was found using Chandra.  
From this, \citet{Antonini2010} suggest that the magnetic activity itself might be highly variable.  We do not detect significant \halpha variability in our high resolution spectra that supports the purported variable magnetic activity; however, we cannot rule out that this is due to the sparse, non-contemporaneous sampling of the optical and X-ray data.  Alternatively, the emission may simply be caused by the continued excitation of HI in the expanding shell of gas and dust by the B3V binary companion, which would indicate that this shell is not completely optically thick.

To ultimately disentangle which scenario is generating the \halpha emission, a simultaneous X-ray and observational campaign is needed.  
As \citet{Antonini2010} note, it is implausible to generate  high X-ray luminosities from the simple interactions of infalling matter with the matter above the stellar photosphere unless extreme accretion rates and infall velocities are involved.  
Thus catching \VMon in the act of producing strong X-ray emission would be a strong indicator that the stellar merger is the ultimate source of any ongoing \halpha activity.  
Even if no X-ray emission is detected, it is certainly worthwhile to continue to monitor the \halpha activity and search for the reappearance of the B3V companion.  
Further evolution (or lack thereof) will put strong contraints on both the material surrounding \VMon and the genesis event that fueled the outbursts.
  
\section{Acknowledgments}
\label{sec:acknow}

This work is supported at The Aerospace Corporation by the Independent Research and Development program. 
SRL acknowledges support from the Michigan Society of Fellows.
MMK acknowledges generous support from the Hubble Fellowship and Carnegie-Princeton Fellowship.


\begin{table*}[!th] 
\caption{Summary of optical and IR spectroscopic and photometric data.}\label{tab:observation_table}
\vspace{-0.15in}
\center{\begin{tabular}{lccr}
\tableline
Date & Observatory & Instrument & Wavelength Coverage/Filters \\
\tableline
2008 Oct 12 & APO  & ARCES  & $3600$-$10000$ $\AA$\\
2009 Nov 01 &      &        &\\
2009 Nov 27 &      &        &\\
2010 Feb 27 &      &        &\\
2010 Mar 25 &      &        &\\
2012 Sep 21 &      &        &\\
\tableline
2008 Oct 12 & APO  & DIS    & blue: $4500$-$5500$ $\AA$ \\
2008 Nov 23 &      &        & red:  $6000$-$9000$ $\AA$\\
2009 Jan 15 &      &        & \\
\tableline
2009 Apr 19 & Gemini South  & T-ReCS & N Lo-Res: 7.7-12.97 $\mu$m\\
2009 Sep 20 &      &        & Si-5: $11.09$-$12.22$ $\mu$m\\
            &      &        & Si-6: $11.74$-$12.92$ $\mu$m\\
            &      &        & Qa:   $17.57$-$19.08$ $\mu$m\\
\tableline
2008 Nov 26 & NASA IRTF     & SpeX   & $0.8$-$2.5$ $\mu$m\\
2009 Jan 10 &      &        &\\
\tableline
2008 Sep 05 & NASA IRTF    & BASS   & $2.9$-$13.5$ $\mu$m\\
2009 Dec 02 &              &        &\\
\tableline
\end{tabular}}
\vspace{0.1in}
\end{table*}
\pagebreak

\begin{table*}[!th]
\caption{Best   fit  parameters  for   two   component  model.}

\vspace{-0.15in}
\begin{center}
\begin{tabular}{lcccccr}
\tableline
&Year&$R_{p}$ (AU)&$T^{p}_{BB}$ (K)&$R_{s}$ (AU)&$T^{s}_{BB}$ (K)&$\epsilon$\\
\tableline
L04 & 2003& 5.6 & 2100 & 28 & 750 & 0.67\\
L07 & 2005& 4.2 & 2100 & 129.25 & 375 &\\
Predictions based on L04 model & 2009 & 4.2 (adopting L7 value) & 2100 & 230.5 & 283 & 0.67\\
Best fit to observed data (current work) & 2009 & 4.3$\pm$2 & 2370$\pm$400 & 263$\pm$10 & 285$\pm$2& 0.65$\pm$.2\\  
\tableline
\end{tabular}
\end{center}
\vspace{0.1in}
\tablecomments{Tabulated values  include the radius of the  primary star ($R_{p}$),
  the blackbody  temperature of  the primary star  ($T^{p}_{BB}$), the
  radius  ($R_{s}$) and  temperature ($T^{s}_{BB}$)  of  the expanding
  dust shell,  and a  uniform attentuation factor  ($\epsilon$).  Note
  that  L04 and  L07 assume  the velocity  of the  expanding  cloud is
  $v_{s}=160$ km  s$^{-1}$ and  the distance to  \VMon is  $D_{V838 Mon}
  \sim 6$ kpc.}\label{table:model}
\end{table*}
\pagebreak

\pagebreak

\pagebreak

\pagebreak

\pagebreak

\pagebreak

\pagebreak

\clearpage
 \newcommand{\noop}[1]{}

\end{document}